\documentclass[journal,comsoc]{IEEEtran}
\usepackage{graphicx}
\usepackage{array}
\usepackage{color}
\usepackage{epsf}
\usepackage{times}
\usepackage{epsfig}
\usepackage{graphicx}
\usepackage{epstopdf}
\usepackage{hyperref}
\usepackage{cite}
\usepackage{amsmath}
\usepackage{amssymb}
\usepackage{amsxtra}
\usepackage{amsthm}
\usepackage{bbm}
\usepackage{algorithmic}
\usepackage{algorithm}
\usepackage{subfigure}
\usepackage{authblk}
\usepackage{url}
\usepackage{comment}
\usepackage{siunitx}
\usepackage{lipsum}
\usepackage{kantlipsum}
\usepackage{dblfloatfix}
\usepackage{adjustbox}
\usepackage{tabularx}
\usepackage{multirow}
\usepackage{float}
\usepackage{framed}
\usepackage{caption}

\begin{document}
\title{Towards Sustainable Internet of Underwater Things: UAV-aided Energy Efficient Wake-up Solutions}
\author{Muhammad Muzzammil, Nour Kouzayha,~\IEEEmembership{Member,~IEEE,} Nasir Saeed, ~\IEEEmembership{Senior Member,~IEEE,} and Tareq Y. Al-Naffouri, ~\IEEEmembership{Senior Member,~IEEE}
\thanks{M. Muzzammil is with the College of Underwater Acoustic Engineering, Harbin Engineering University, China, Email: muzzammilm@hrbeu.edu. \newline
N. Kouzayha and T. Y. Al-Naffouri are  with the Division of Computer, Electrical and Mathematical Sciences and Engineering, King Abdullah University of Science and Technology, Thuwal 23955-6900, Saudi Arabia  Email: \{nour.kouzayha, tareq.alnaffouri\}@kaust.edu.sa.\newline 
N. Saeed is with the Remote Sensing Unit, Department of Electrical Engineering, Northern Border University Arar, Saudi Arabia. Email: mr.nasir.saeed@ieee.org.}}

\maketitle
	\begin{abstract}
	With the advancements in underwater wireless communications, internet of underwater things (IoUT) realization is inevitable to enable many practical applications, such as exploring ocean resources, ocean monitoring, underwater navigation, and surveillance. The IoUT network comprises battery-operated sensor nodes, and replacing or charging such batteries is challenging due to the harsh ocean environment. Hence, an energy-efficient IoUT network development becomes vital to improve the network lifetime. Therefore, this paper proposes unmanned aerial vehicle (UAV)-aided energy-efficient wake-up designs to activate the underwater IoT nodes on-demand and reduce their energy consumption. Specifically, the UAV communicates with water surface nodes, i.e., buoys, to send wake-up signals to activate the IoUT sensor nodes from sleep mode. We present three different technologies to enable underwater wake-up: acoustic, optical, and magnetic induction-based solutions. Moreover, we verify the significance of each technology through simulations using the performance metrics of received power and lifetime. Also, the results of the proposed on-demand wake-up approach are compared to conventional duty cycling, showing the superior performance of the proposed schemes. Finally, we present some exciting research challenges and future directions.
	
	\end{abstract}
	
	\IEEEpeerreviewmaketitle

\section{Introduction}
\label{sec_intro}
Water in the form of oceans, seas, and rivers covers two-third of the planet's surface, which is yet mostly unexplored ~\cite{khalil2020towards}. With $90$\% international trade through maritime transportation and about $500$ million people's food from the coral reefs~\cite{WHOI:2022:Online}, the vast oceanic system exploration through underwater research and development will have a significant impact on human life. Due to recent technological advancements in the development of underwater wireless communication networks (UWCNs), the practical implementations of the Internet of Underwater Things (IoUT) can be realized~\cite{khalil2020towards}. IoUT offers an opportunity to explore oceanic systems and provides numerous scientific, industrial, and military applications. 

Unlike the terrestrial Internet of Things (IoT), where radio frequency (RF) signals are primarily used for communication between IoT nodes, the harsh underwater environment poses significant challenges for RF-based communication due to substantial signal attenuation and absorption in the aquatic environment. Therefore, researchers have widely adopted underwater acoustic communication (UAC) technology to communicate over longer distances (up to several kilometers). Nevertheless, the acoustic technology faces challenges of limited bandwidth, low data rate, and high power consumption~\cite{stojanovic2016acoustic}. Therefore, the underwater research community investigated alternative technologies, such as optical and magnetic induction (MI). The optical-based systems can be a suitable choice for medium-range underwater applications, offering a high data rate, low link delay, low cost, and low energy consumption~\cite{NasirUWOCSurvey19}. However, optical signals require line of sight communication, are affected by water turbidity and are absorbed and scattered quickly in the underwater environment. Therefore, researchers further investigated the MI technology, which proved to be a reliable, secure, and low energy option for medium-range applications by offering a reasonable data rate in Mbps and lower link delay~\cite{muzzammil2020fundamentals}. Another unique feature of the MI technology is providing smooth inter-medium communication (air-to-water or water-to-air) due to the same magnetic permeability in air and water, while acoustic and optical technologies may fail to work efficiently in such a scenario.

Besides the harsh underwater wireless communication channel challenge, another big issue for IoUT networks is optimizing their energy resources. Since most of the underwater devices in the IoUT network are mainly battery-operated, the traditional operating ways of charging or replacing these batteries are costly and non-scalable. Hence, we need to develop an energy-efficient IoUT system that minimizes the overall energy consumption of the network. Towards this end, researchers have developed low power radios and adopted duty cycling-based medium access control (MAC) approaches to improve the network energy efficiency by periodically putting the sensor nodes in sleep/active mode~\cite{piyare2017ultra}. However, challenges of idle listening, overhearing, latency, time synchronization overhead, and continuous or multiple transmission (in the case of asynchronous operation) are associated with duty cycling optimization approaches. Furthermore, the power consumption of listening and transmission modes is similar in low-power radios. Therefore, these low-power radios and duty cycling MAC approaches may not significantly contribute to the overall network energy consumption minimization.

Consequently, the research community started investigating alternate methods, such as designing wake-up radios (WuRs), to overcome the limitations of low power radios and duty cycling approaches. The WuR is an ultra-low-power additional module that consumes several orders of magnitude lower power than the primary low power radios. Hence, the WuR module is always on while the rest of the node's circuitry is in sleep/off mode until needed. In WuR, the wake-up transmitter (WuTx) sends a modulated wake-up signal (WuS), which upon being received by the wake-up receiver (WuRx), triggers the main micro-controller of the node to activate it for data transmission/reception. There are three modes to operate WuR: active (operates on battery power), passive (harvesting energy directly from WuS), and hybrid (few WuR components run on battery while others harvest energy from WuS). 
\begin{figure}
    \centering
    \includegraphics[width=0.5\textwidth]{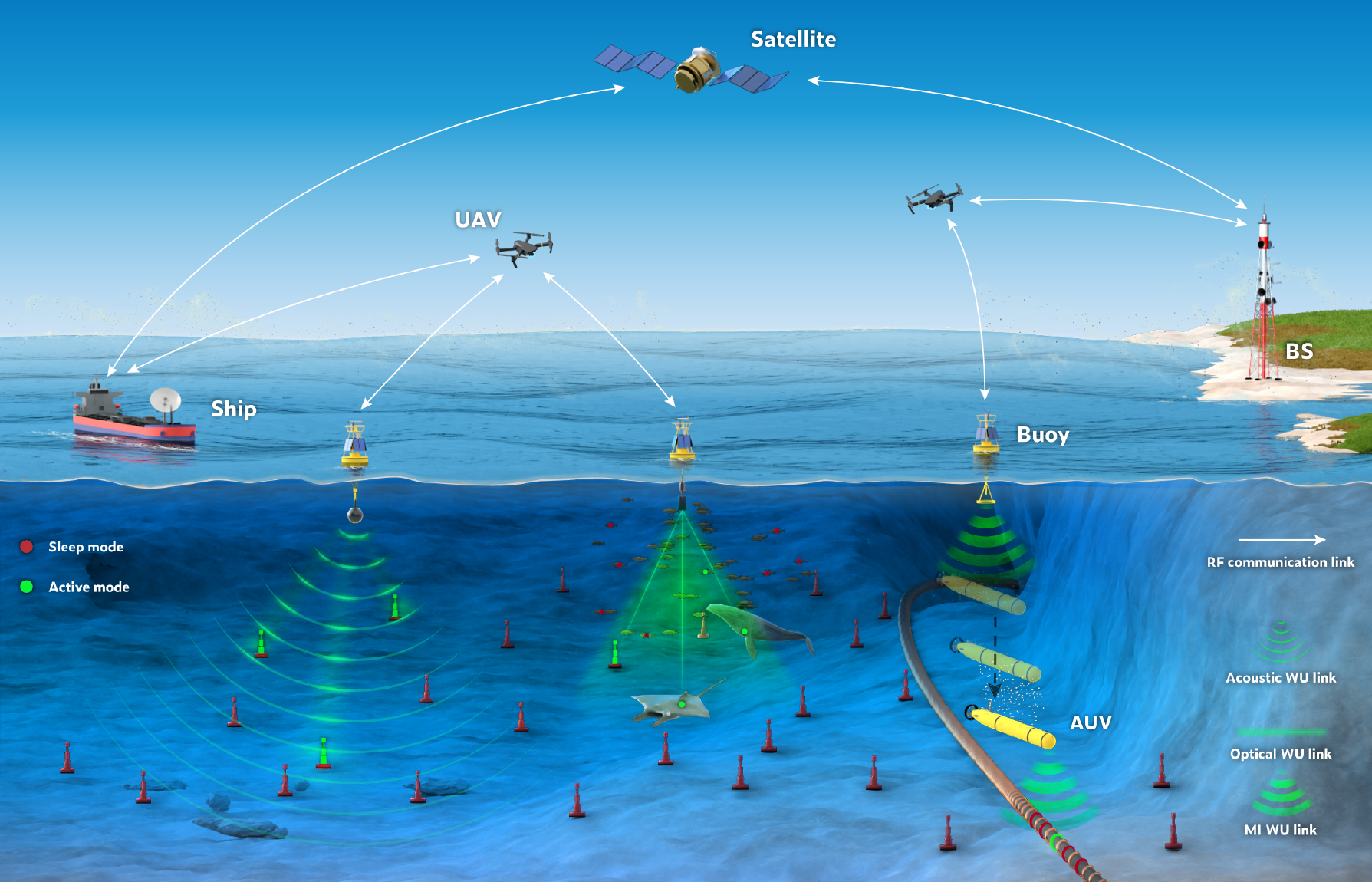} 
    \caption{A generic IoUT network with the different proposed wake-up schemes.}
    \label{fig:generic_IoUT}
    \vspace{-0.2cm}
    \end{figure} 
The wide utilization of this energy-efficient WuR approach in the terrestrial IoT networks motivates us to adapt it to the energy-scarce underwater environment. Therefore, this paper presents energy-efficient on-demand wake-up-based design solutions using acoustic, optical, and MI technologies to activate IoUT nodes. The proposed solutions also utilize UAVs that can be beneficial in many scenarios, such as disastrous situations or activation of a specific underwater IoT node where it may not be possible for the surface node to activate it due to WuS transmission power limitations. Moreover,
whether acoustic, optical, or MI, each technology has its design challenges. Therefore, this paper further studies various strategies adopted in each technology and explores their applicability for underwater applications based on different performance metrics such as power consumption, receiver sensitivity, communication range, data rate, and mobility. To summarize, the following are the main contributions of this paper:
\begin{enumerate}
    \item First, the opportunities and limitations of a UAV-aided IoUT system are explored and the concept of on-demand wake-up for IoUT nodes is introduced to minimize the energy consumption of the network. 
    \item Then, we propose three different conceptual wake-up designs to implement the UAV-aided IoUT system, including RF-acoustic, RF-optical and RF-MI which use the RF technology in the air while using acoustic, optical and MI technologies in underwater environment. 
    \item After that, we conduct simulations to analyze the behavior of the IoUT nodes for each wake-up design solution. Specifically, the performance is evaluated in terms of the energy consumption, sensitivity, range, and lifetime for each approach. Based on the obtained results, policies are defined for the use of UAVs and the choice of the best technology to wake-up IoUT nodes.
    \item Finally, we present research challenges and future directions to fill the gap in literature and expedite the implementation of the proposed schemes for practical IoUT applications.
\end{enumerate}

 
\section{IoUT Nodes Wake-up Solutions}
This section presents three UAV-aided energy-efficient on-demand wake-up solutions for IoUT node activation: RF-acoustic, RF-optical, and RF-MI. Fig.~\ref{fig:generic_IoUT} illustrates the generic IoUT architecture with the three wake-up strategies where the RF link is used for communication between UAV and surface node in the first phase, and acoustic, optical, or MI wake-up links are used for IoUT nodes activation in the second phase. In the following, we discuss both of these stages separately.
\begin{figure*}[]
    \centering
    \vspace{-0.50cm}
    \includegraphics[width=0.95\textwidth]{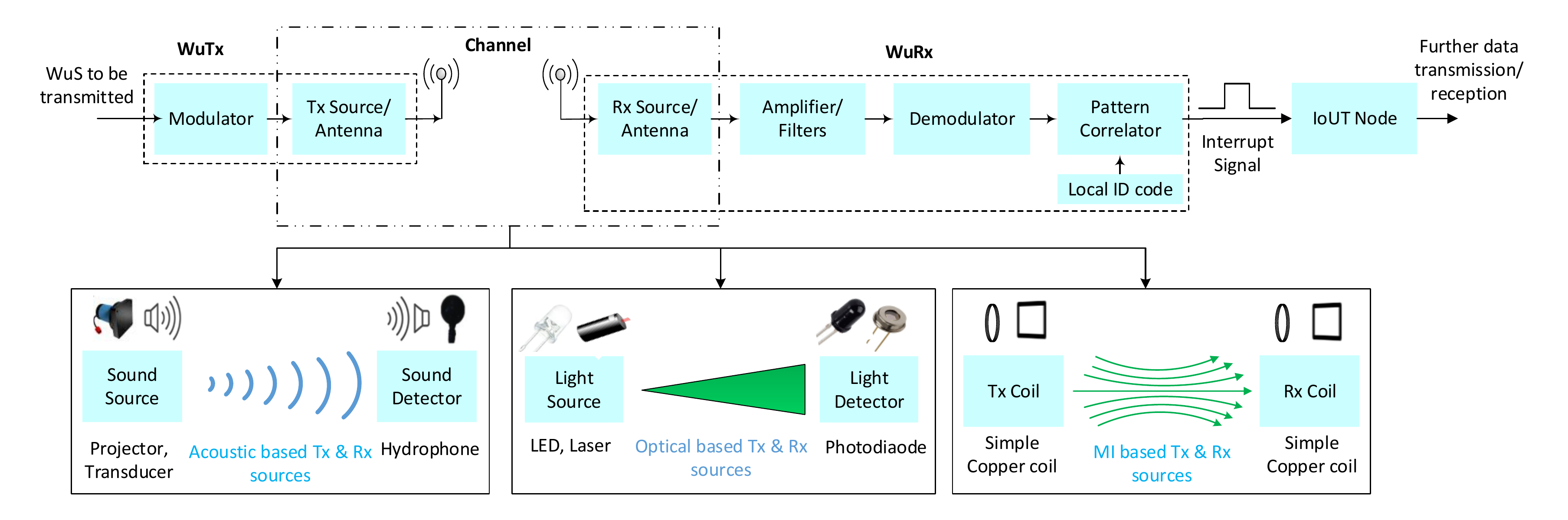} 
    \caption{Block diagram of wake-up transceivers: Acoustic, optical, and MI.}
    \label{fig:Wu-transceivers}
        \vspace{-0.3cm}
    \end{figure*}
\subsection{RF-based UAV to Surface Node Communication Link}\label{subsec:rf}
In the first stage, the UAV transmits an RF-based information signal to the water surface node, i.e., buoy, with the message of activating the IoUT node through a wake-up approach. Once the buoy receives this message, it transmits the WuS towards the IoUT node. Besides buoys, using autonomous underwater vehicles (AUVs) for an air-to-water surface link is a better alternative since the AUVs can reach out to the underwater IoUT nodes directly. 

The buoy node performs an essential job of first receiving an RF-based information signal from the UAV and then transmitting either acoustic-, optical-, or MI-based WuS towards the IoUT node. Also, the buoy can harvest energy directly from sunlight through the solar panel to enhance its energy efficiency, as shown in Fig.~\ref{fig:generic_IoUT}. Furthermore, the buoy can directly harvest energy from the incoming RF signal. Because of energy harvesting capabilities, the buoy will have sufficient energy and therefore can transmit WuS at a much higher power towards the IoUT node, which leads to an extended range and larger area coverage. Alternatively, if the buoy has limited energy, the on-demand wake-up strategy can also be used for the UAV-buoy link. In this case, a UAV first transmits an RF-based WuS towards the buoy for its activation, and then the buoy will further send acoustic-, optical-, and MI-based WuS towards IoUT nodes for activation. Hence, enough energy can be saved at the buoy node and will lead to an increased network lifetime. However, in both cases: with or without a wake-up approach in the UAV-buoy link, additional hardware components need to be added to the buoy, which leads to an increased cost and size. Note that the UAV-buoy link is RF-based, and therefore the techniques used in terrestrial IoT networks for on-demand wake-up are well applicable. Furthermore, challenges such as UAVs trajectory and buoy placement will also play a significant role in energy and resources optimization.
Once the buoy receives the requirement of WuS from the UAV, it uses the acoustic, optical, and MI-based underwater IoUT nodes' wake-up solutions, where an extra wake-up transceiver is added to the IoUT node to receive and process the WuS transmitted from the surface node. The general block diagram of the added wake-up transceiver module along with specific TX/RX for acoustic, optical, and MI modules are shown in Fig.~\ref{fig:Wu-transceivers}. The following sections explicitly describe each of the IoUT node wake-up solutions.
\vspace{-0.3cm}
\subsection{Acoustic-based IoUT Node Wake-up}\label{subsec_acoustic}
Unlike the wide utilization of RF-based wake-up in terrestrial networks, research on acoustic-based wake-up is still limited, especially in the underwater environment. A few studies focused on the design of WuS in acoustic-based wake-up, or acoustic-based energy harvesting~\cite{sun2019design, guida2020underwater}. 
However, these studies don't address underwater-related challenges and joint aerial to underwater node activation. Although the idea of an acoustic-based wake-up approach seems very simple, the harsh underwater environment and limitations of acoustic technology cause significant design and implementation challenges, such as high attenuation and absorption due to the conductive nature of seawater. Moreover, due to the slow propagation speed of sound waves underwater ($1500$~m/s), latency is a major issue for using acoustic technology for timely on-demand wake-up. 

Receiver sensitivity, cost, and size are other essential factors in an acoustic-based wake-up approach. With low strength WuS reception at WuRx, increasing or decreasing receiver sensitivity can significantly impact the power consumption. In our proposed joint RF-acoustic approach, the surface node needs to be equipped with both an RF module and an acoustic wake-up module, leading to an increased cost and size. Additionally, mobility can be a dominant factor in the on-demand acoustic-based wake-up design as IoUT nodes may experience movement due to water waves and tides, leading to severe multi-path and Doppler effects.

In the acoustic-based wake-up scheme, the surface node transmits on-demand acoustic WuSs to power and activate the IoUT node. The WuS is generated using the sound source (projector or transducers), which is received by a sound detector (hydrophone) at the receiver side and processed for amplification and filtering as shown in Fig.~\ref{fig:Wu-transceivers}. Then, the filtered received WuS is demodulated and its address is compared with the local ID. When the address of WuS is matched, an interrupt is generated to activate the main MCU of the IoUT node. In case of address mismatch, the received WuS is not further processed, and no interrupt is generated; hence, the IoUT node remains in sleep mode. Consequently, a considerable amount of energy can be saved with an on-demand wake-up approach, leading to an increased IoUT node lifetime.

 \begin{figure*}[]
 \vspace{-0.50cm}
 \centering
    \subfigure[\label{subfig:recpower_acoustic}]{\includegraphics[width=0.39\textwidth]{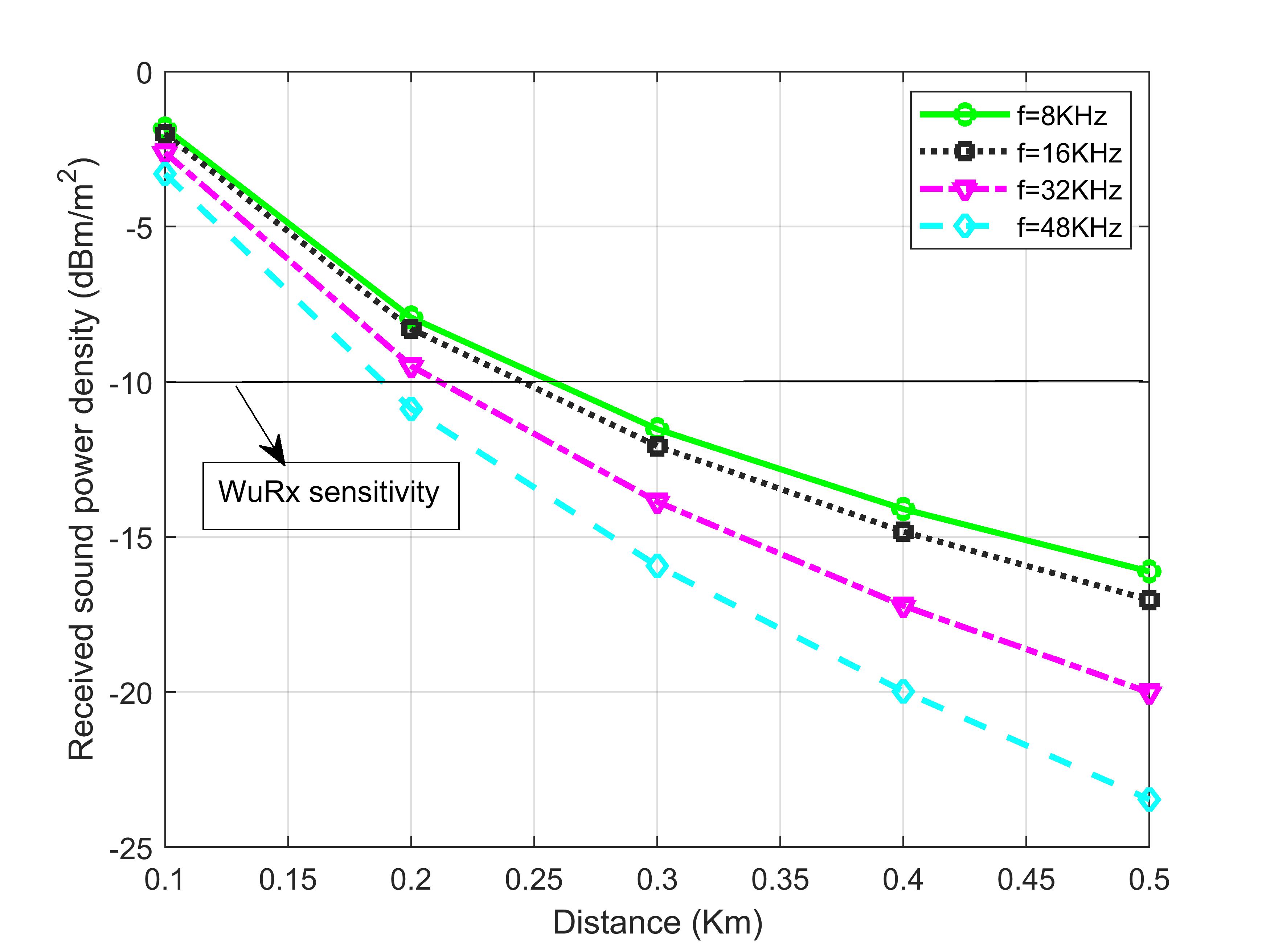}}
  \subfigure[\label{subfig:lifetime-acoustic}]{\includegraphics[width=0.39\textwidth]{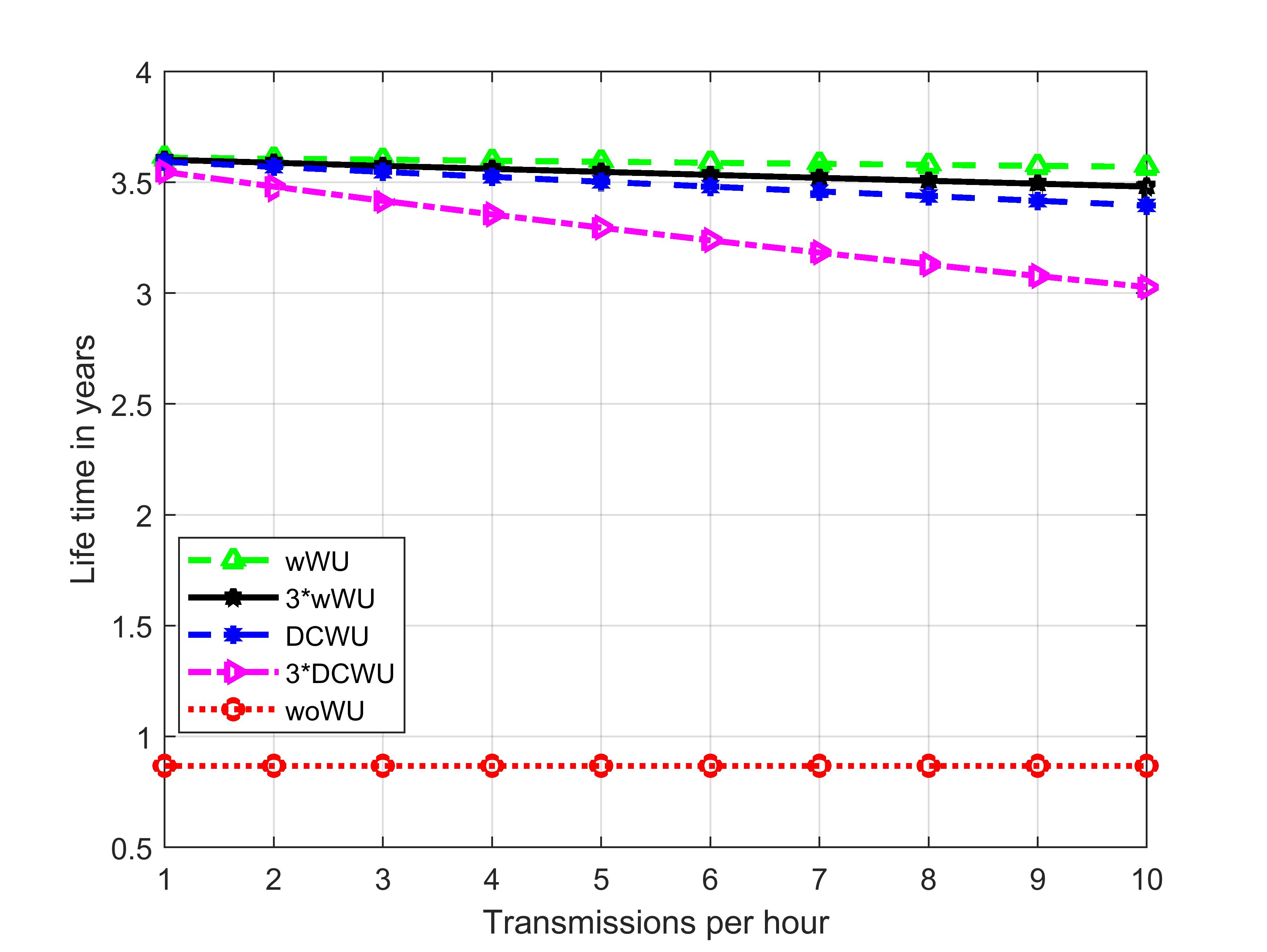}}
  \vspace{-0.3cm}
 \caption{Acoustic-based IoUT nodes wake-up: (\textbf{a}) received power vs. distance, and (\textbf{b}) life time vs. transmissions per hour.}
 \label{fig:acoustic1}
 \end{figure*}
   
To support the above discussion, we carried out simulations of received sound power density vs. distance and IoUT node lifetime vs. the number of transmissions per hour, as depicted in Fig.~\ref{fig:acoustic1}. We consider the ocean density = $1000$~Kg/$\text{m}^3$, sound speed = $1500$~m/s, sound source level = $190$~dB, and receiver sensitivity of $-10$~dBm~\cite{lattanzi2013sub}. From Fig.~\ref{subfig:recpower_acoustic}, it is clear that the received sound power density decreases as the distance increases. Also, the received sound power density is much higher when using a lower frequency ($8$~KHz case) compared with a higher frequency ($48$~KHz case). 
With the WuRx sensitivity of $-10$~dBm, acoustic-based WuS can reach over a distance of about $260$~m for a frequency of $8$~KHz vs. $190$~m for the $48$~KHz case.

The lifetime of an IoUT node based on acoustic technology is evaluated by having modes of no wake-up, duty cycling (DC) based wake-up, and on-demand wake-up. We consider the battery capacity of an IoUT node to be $950$~mAH, and the node takes $1$~sec of time for active transmission and is in sleep mode for the rest of the time. The current consumption in active and sleep modes are $500$~$\mu$A and $15$~$\mu$A, respectively~\cite{bannoura2016acoustic}. From Fig.~\ref{subfig:lifetime-acoustic}, we can observe a significant difference between the IoUT node lifetime with on-demand wake-up (wWU) and without wake-up (woWU) because the IoUT node is always in the active mode in case of the no wake-up approach and therefore consumes the same amount of current continuously from the battery, hence, the lifetime is low. However, in the case of duty cycling-based wake-up (DCWU) and on-demand wake-up, the IoUT node's current consumption depends on the number of active transmissions in a given time (per hour/day). The number of active transmissions in duty cycling-based wake-up is higher than the on-demand wake-up approach due to the pre-defined schedule and the infrequent transmissions characteristic of IoUT nodes. For instance, we characterize the on-demand wake-up approach by the requirement of IoUT node activation of $1$~time per hour; however, in the duty cycling approach, the transmission rate is set to $5$~times per hour, which means that on-demand-based wake-up consumes four-times less current than duty cycling, resulting in a higher lifetime. 
These results suggest that the proposed on-demand acoustic wake-up approach is a better alternative compared to the high cost and non-scalable conventional battery recharge approaches for the lifetime enhancement of IoUT networks, especially in harsh underwater environments.
 
\subsection{Optical-based Underwater Node Wake-up}\label{subsec_opto}
Unlike the acoustic counterpart, research on optical-based wake-up is quite limited, and only a few recent works have investigated it for underwater applications. For instance, the authors in~\cite{de2020toward} introduce simultaneous light-wave information and power transfer (SLIPT) technology for harvesting light waves to power underwater IoT nodes and transfer data at the same time; however, this study lacks the use of a wake-up approach for further improving the network lifetime. Due to its low power consumption compared to the acoustic technology, the optical-based wake-up approach can be beneficial in terms of enhanced network lifetime. 
Nonetheless, an optical-based wake-up approach also has design challenges. For example, the optical-based WuS leads to a high data rate and lower latency but for a limited communication range.

\begin{figure*}[t]
 \vspace{-0.50cm}
 \centering
    \subfigure[\label{subfig:recpower_optical}]{\includegraphics[width=0.39\textwidth]{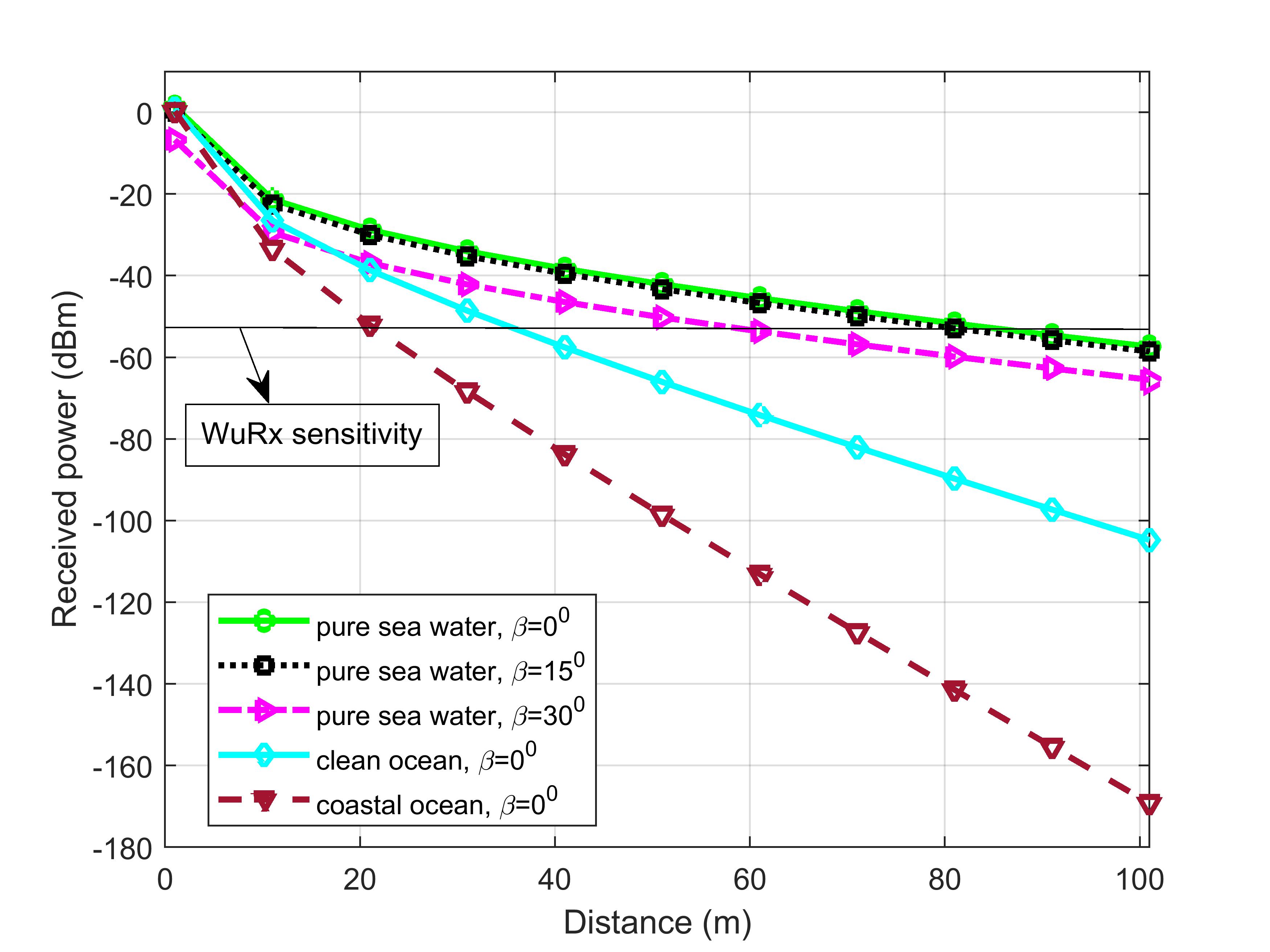}}
  \subfigure[\label{subfig:lifetime-optical}]{\includegraphics[width=0.39\textwidth]{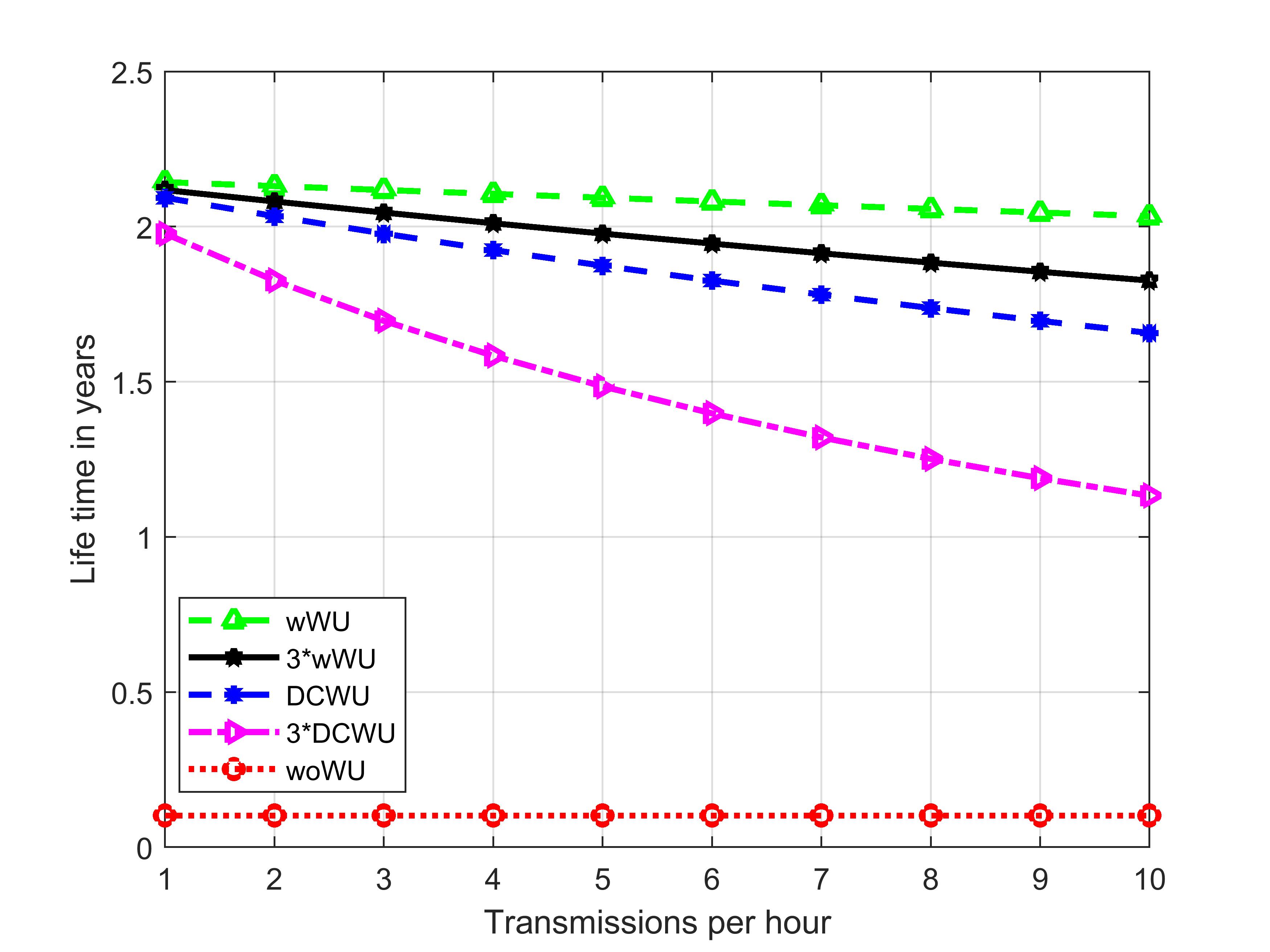}} 
  \vspace{-0.2cm}
 \caption{Optical-based IoUT nodes wake-up: (\textbf{a}) received power vs. distance, and (\textbf{b}) life time vs. transmissions per hour.}
 \label{fig:optical}
 \end{figure*}
In optical-based underwater IoT node wake-up, the surface node or AUV transmits optical WuS towards the node to power and activate it for data transmission/reception. The working operation of optical based wake-up is similar to acoustic-based wake-up as described in subsection~\ref{subsec_acoustic}. However, a light source such as LED or laser is used for WuS transmission and a photo diode is used as a light detection source for detecting the WuS at the receiver side as shown in Fig.~\ref{fig:Wu-transceivers}. Parameters such as the optical wake-up module's receiver sensitivity, cost, size (LEDs, laser, and photo-diodes), and the type of underwater optical channel affect the optical wake-up. The optical technology also requires line of sight communication and is more sensitive to mobility and misalignment between the transmitter and the receiver. Fig.~\ref{fig:optical} shows the impact of these parameters on the received optical power versus distance, different types of water, and varying misalignment angle $\beta$ where the simulation parameters are set to: transmit power~$=250$~mW, transmit and receiver aperture area~$=0.0011$~$m^2$, transmitter beam divergence angle~$=0.5^0$. Note that the received power decreases with distance at a much higher rate than the acoustic technology because the attenuation and absorption of an optical signal are quite significant, leading to shorter ranges. By considering a practical value of  WuRx sensitivity as $-53$~dBm~\cite{mathews2010low}, the successful wake-up call can reach over a distance of about $90$~m in the case of perfect alignment (see Fig.~\ref{subfig:recpower_optical}).  

We also evaluate the lifetime of the proposed optical-based IoUT node wake-up concerning the number of active transmissions in Fig.~\ref{subfig:lifetime-optical}. For the simulations, we consider the battery capacity of $950$~mAH, active transmissions of $1$~per~sec, current consumption of $3.6$~mA in the active mode, and $83$~$\mu$A in the sleep mode~\cite{mathews2010low}. A similar lifetime trend can be depicted in Fig.~\ref{subfig:lifetime-optical}, where the lifetime of an IoUT node without a wake-up approach is much lower as compared to DC-based and on-demand wake-up. Here, on-demand wake-up outperforms other methods due to the small number of active transmissions per hour, consuming less power. To summarize, the on-demand optical-based wake-up approach can be useful for short-range, delay-sensitive, and energy-scarce underwater applications.

\subsection{MI-based Underwater Node Wake-up}\label{subsec_MI}
The MI communication technology enjoys several advantages over acoustic and optical technologies by offering a high data rate and low link delay due to magnetic field stability, predictable channel response, and no multipath. In addition, MI technology is less power-hungry and more efficient in energy harvesting than acoustic and optical technologies.
Fig.~\ref{fig:Wu-transceivers} shows the block diagram of MI-based wake-up design where coils work as transceiver antennas for transmission and reception of WuS.

Although energy harvesting through MI is previously studied in the underwater environment, the wake-up approach is rarely investigated, and only a few MI-based wake-up studies exist~\cite{ahmed2018design}. 
Like the acoustic and optical technologies, the passive MI-based wake-up also exhibits design challenges, such as selecting the coupling type (inductive coupling or resonant coupling) and misalignment between the coils. Inductive coupling works well for near-field communication, while resonant coupling performs better for long transmission ranges and is mostly adopted in underwater environments. In underwater MI communication, the magnetic field signal power rapidly attenuates, i.e., at $1/d^6$, where $d$ is the communication range. Therefore, using it for low-power WuR makes it even more range-limited. However, wake-up latency would be minimal compared to acoustic and optical because of the instant creation/reception of magnetic field-based WuS.
\begin{figure*}[t]
 \vspace{-0.50cm}
 \centering
    \subfigure[\label{subfig:recpower_MI}]{\includegraphics[width=0.39\textwidth]{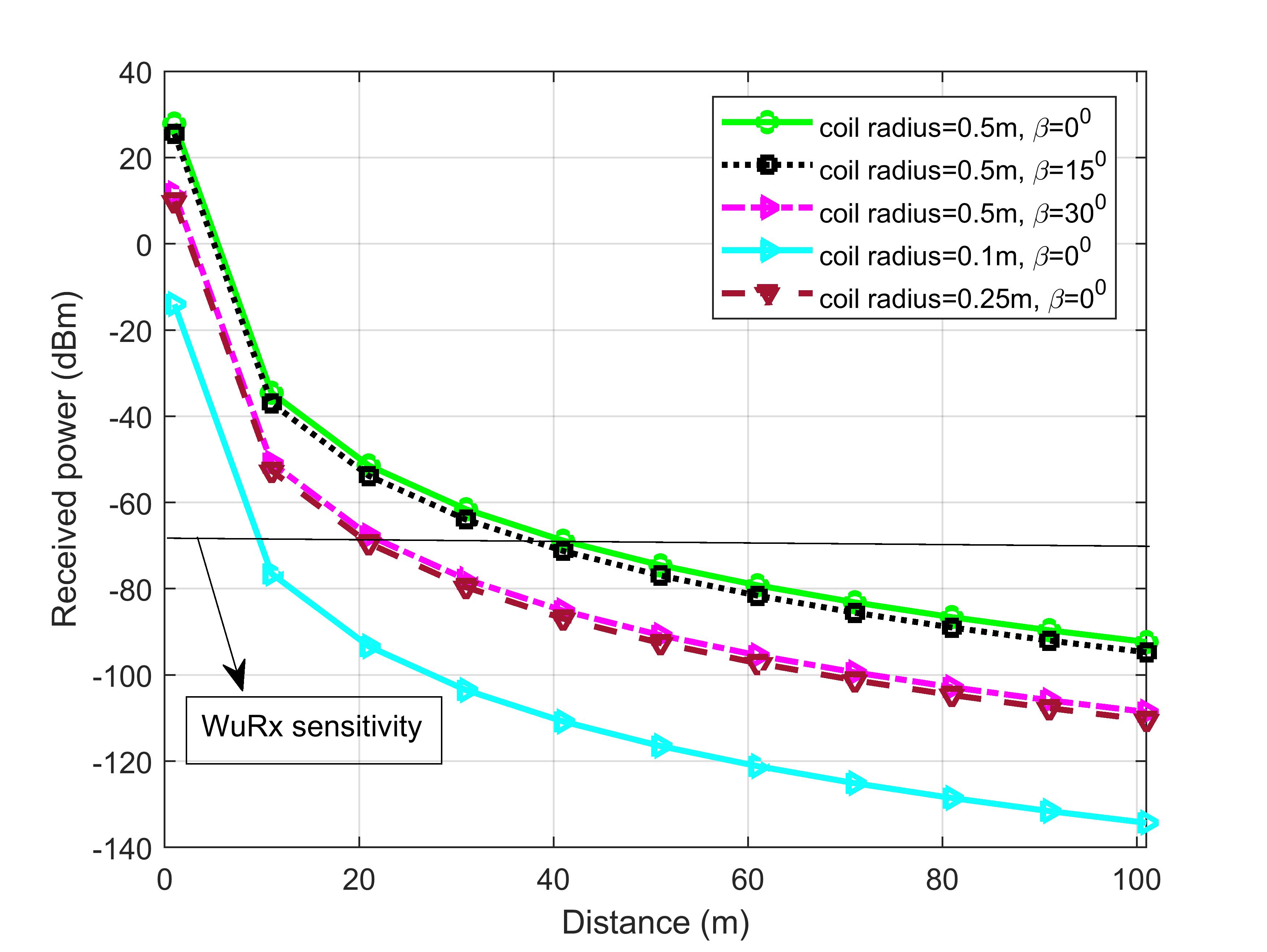}} 
  \subfigure[\label{subfig:lifetime-MI}]{\includegraphics[width=0.39\textwidth]{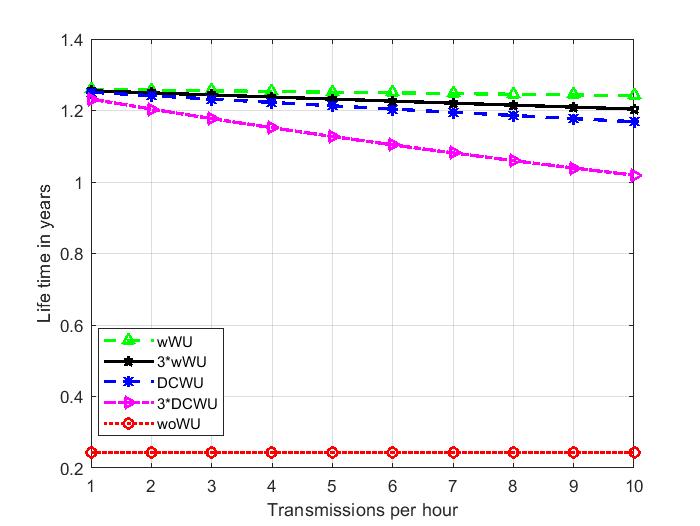}}
 \caption{MI-based IoUT nodes wake-up: (\textbf{a}) received power vs. distance, and (\textbf{b}) life time vs. transmissions per hour.}
 \label{fig:acoustic}
 \end{figure*}
In MI-based wake-up, the receiver sensitivity needs to be carefully adjusted, as perfect alignment between WuTx and WuRx coils is mandatory for high power reception. Fig.~\ref{subfig:recpower_MI} shows the MI field received power versus distance under different coil radii and alignment angles $\beta$. The simulation parameters are: operating frequency~$=75$~KHz, magnetic permeability constant $\mu=4\pi \times 10^{-7}$~H/m, and the number of turns in transmit and receive coils is $30$. The figure shows a much faster-decreasing trend of received power vs. distance as compared to acoustic and optical technologies. Here, when the MI WuRx sensitivity is $-69$~dBm~\cite{ahmed2018design}, the successful wake-up call can last over a distance of about $44$~m with a perfectly aligned transmitter and receiver coils and reduces when misalignment increases, as shown in Fig.~\ref{subfig:recpower_MI}. 

We perform the lifetime analysis of an IoUT node with MI technology with the same battery capacity ($950$~mAH) and active transmission time of $1$~sec. The current consumption in active mode is $0.49$~mA and in sleep mode is $43$~$\mu$A based on the MI node design~\cite{ahmed2018design}. Fig.~\ref{subfig:lifetime-MI} shows a similar lifetime trend for without wake-up, DC-based wake-up, and on-demand wake-up approaches. MI-based wake-up solution can therefore be proved beneficial in medium range, delay-sensitive harsh underwater environment due to relatively better underwater channel performance over acoustic and optical technologies. 

 \vspace{-0.2cm}
\subsection{Comparison of Proposed Wake-up Solutions}
\label{subsec:comp}
This section provides a comparative analysis of the proposed wake-up solutions relating to real-time underwater application scenarios, as summarized in Table~\ref{tab:comp}. 
\begin{table*}[]
\centering
\caption{Comparison of underwater IoUT wake-up technologies.}
\label{tab:comp}
\begin{tabularx}{\textwidth}{|X|X|X|X|}
\hline
\textbf{Features} & 
                                       \textbf{Acoustic Wake-up} & \textbf{Optical Wake-up} & \textbf{MI Wake-up} \\
\hline     
Speed           & Underwater sound speed, i.e., $1500m/s$ & Speed of light, i.e., $3 \times 10^8m/s$ & Speed of light, i.e., $3 \times 10^8m/s$ \\
\hline
Range   & Long range ($ \approx$Km)  & Medium range ($\approx$10-100m) & Medium range ($\approx$10-100m) \\
\hline
Data rate   & Kbps  & Gbps  & Mbps \\
\hline
Frequency band  & Few Kilohertz ($\approx 8-48$Khz) & Gigahertz ($\approx 5 \times 10^{14}$Hz)   & Kilohertz to few Hundred Kilohertz 
\\
\hline
Signal operation    & Audible   & Visible   & Non-audible and non-visible \\
\hline
Channel characteristics & Temperature, salinity, multipaths, and Doppler spread, etc. & Attenuation, scattering, sea noise, and orientation sensitivity etc. & Conductivity and orientation sensitivity etc. \\
\hline
Energy efficiency   & Low      & Average     & High \\
\hline
Receiver sensitivity    & $-10$~dBm~\cite{lattanzi2013sub} & $-53$~dBm~\cite{mathews2010low} & $-69$~dBm~\cite{ahmed2018design} \\
\hline
Cost  & High  & Medium  & Low \\
\hline
\end{tabularx}
\end{table*}
Each of the three proposed wake-up designs has unique features and can be used for specific applications. For example, the acoustic-based wake-up design can be a viable option in case of delay-insensitive applications such as deep-sea exploration/monitoring and transmitting wake-up signals over long distances to activate deep-sea underwater nodes. However, the harsh underwater acoustic channel needs to be dealt as it depends on various parameters such as water temperature, pressure, salinity, multipaths and Doppler spread etc. Further, the cost of acoustic-based equipment is also very high. On the other hand, optical-based wake-up design can be helpful for low-cost and delay-sensitive medium range real-time underwater applications such as surveillance and disastrous monitoring. Beside these merits, factors such as attenuation, scattering, sea noise and orientation sensitivity greatly affect the performance of optical systems. Similarly, MI-based wake-up design can perform instant transmission of wake-up signal; therefore, it can also be proved significant in secure, low-cost, delay-sensitive, and medium-range applications such as underwater oil and gas pipeline leak detection. However, the MI based wake-up performance is mainly affected by conductivity and orientation sensitivity that needs to be carefully considered.
In addition, it is worth noting that one can use different technologies for wake-up signal and data transmission for a specific underwater application; however, the cost and complexity of the network will increase due to additional hardware modules. 
\section{Research Challenges and Future Directions} \label{sec:future_direction}
This section highlights some exciting research challenges and future directions related to IoUT wake-up designs, such as inter-medium wake-up, trajectory and resource optimization, wake-up range extension, energy harvesting methods, and novel wake-up protocol designs. In the following, we discuss each of these future research directions.
\vspace{-0.2cm}
\subsection{Inter-medium Wake-up}\label{subsec:intermediumWU}
Terrestrial networks mostly use RF technology, while underwater environments primarily utilize acoustic technology. However, due to their own limitations, these technologies cannot be used as a stand-alone technology for inter-medium communication. For inter-medium communication, air-to-water or water-to-air, between aerial and underwater nodes is conventionally realized using a surface node equipped with dual technology modems. However, the recent development of direct inter-medium communication using optical and MI technologies can be a progressive solution for underwater IoUT nodes' wake-up. In an optical band, blue/green laser light can penetrate the air-water surface, while in MI, the magnetic field has the unique feature of inter-medium communication due to the similar magnetic permeability of air and water. Therefore, in our proposed wake-up approaches, UAVs can either be equipped with optical or MI modules that can directly transmit WuS towards the underwater IoT node without using a buoy. Advantages of this direct inter-medium communication include low latency, low cost, high data rate, reduced overall energy consumption, and enhanced network lifetime. However, both technologies face a number of challenges, such as orientation sensitivity and limited communication range.

Another exciting direction to solve the inter-medium communication and wake-up challenge can be the development of floating gateways (UAVs capable of landing on the water surface)  equipped with either acoustic, optical, or MI technologies to transmit on-demand wake-up signals for underwater IoT node activation. These floating gateways can efficiently use the hardware resources, reducing the overall cost and energy of the IoUT network. Nevertheless, development of UAVs working as floating gateways for practical inter-medium wake-up is still an open research problem.
\vspace{-0.2cm}
\subsection{UAVs Trajectory and Resource Optimization}
\label{subsec:uavtraj}
The IoUT network consists of various nodes such as UAVs, buoys, AUVs, and underwater sensor nodes. Among these nodes, UAVs/AUVs are promising technologies to improve different aspects of IoUT networks, such as increased coverage, mobility support, and resource optimization. Since our focus is to minimize the energy of an IoUT network,  UAVs/AUVs trajectories optimization can play a significant role in minimizing the overall power consumption. As in our proposed on-demand wake-up designs, UAVs initiate the wake-up process by transmitting WuS signals to the buoys; hence using an optimized UAVs trajectory can increase energy efficiency. Similarly, the optimal number and placement of buoys and underwater sensor nodes can also lead to an efficient resource utilization that can save the cost and energy of the overall IoUT network. The research community can pursue these trajectory and resource optimization directions to develop an energy-efficient IoUT network for practical implementations.
\vspace{-0.30cm}
\subsection{Range Extension through RIS}\label{subsec:range}
As discussed earlier, a trade-off exists between transmission range and data rate in an underwater medium while using acoustic, optical, and MI technologies transmitting wake-up signals. Recently, reconfigurable intelligent surfaces (RIS) are emerging as a range-extension technology to increase the communication range and data rate in the underwater medium. The RIS can also help extend the transmission range of WuS by using the idea of intelligent reflections for acoustic, optical, and MI technologies. Few recent works developed acoustic and optical based RIS for range extension in the underwater medium~\cite{sun2021acoustic, ramavath2022evaluation}; however, acoustic-, optical-based RIS related to wake-up signal range extension is unavailable. Similarly, MI-based RIS are yet not explored and need further investigations.
Moreover, various other challenges are associated with RIS development for underwater mediums, such as hardware design, beamforming capabilities, mobility modeling, and network coverage and operation analysis. 
\vspace{-0.1cm}
\subsection{Passive Underwater Node Wake-up}\label{subsec:passiveWU}
In passive wake-up schemes, the IoUT nodes do not use the battery power for wake-up and instead harvest energy from the incoming WuS. Once the targeted IoUT node harvests enough energy, a wake-up trigger is generated. This approach minimizes the energy consumption and enhance the network lifetime. However, several challenges are associated with this approach. For example, we need to ensure enough power transmission from WuTx to harvest energy for initializing trigger operation and significant latency as the node will first harvest energy and then wake-up the node. Additionally, receiver sensitivity and mobility of underwater IoT nodes critically affect the performance of the passive wake-up approach. This is especially important for the MI and optical transmissions because nodes are more prone to misalignment due to water waves and tides.
\vspace{-0.15cm}
\subsection{Novel MAC Protocols Design}\label{subsec:protocols}
The network topology, channel conditions, routing, and MAC protocols play an essential role in defining the overall power consumption of the network. While designing the MAC protocols, channel sharing also needs to be carefully considered in the case of an in-band system where the wake-up module and main sensor module operate at the same frequency bands. Furthermore, the wake-up module and main sensor nodes module have different communication ranges, leading to an asymmetric network, which is quite challenging for the wake-up-based protocol design. This asymmetry is further exacerbated by the underwater channel variations and node mobility, so a robust adaptive MAC protocol is needed.
Moreover, cognitive radio-based wake-up modules are suggested to overcome the channel variability issue. Most of these challenges still need detailed investigations, explicitly considering the harsh underwater channel environment. 
 
\section{Conclusions and Outlook}\label{sec:conclusion}
This paper presents three different on-demand wake-up design solutions: RF-acoustic, RF-optical, and RF-magnetic induction to activate the underwater IoUT nodes. The wake-up process is performed in two stages: 1) the on-demand wake-up call is initiated from the air through UAV by sending an RF-based information signal to the buoy node, and 2) the buoy node further transmits WuS based on acoustic, optical, or MI technology towards the IoUT node. We performed simulations to evaluate the performance of the proposed on-demand wake-up solutions to verify the significance of each technology. Simulations show that the on-demand wake-up significantly improves the lifetime of the IoUT network compared to the without wake-up and duty cycling-based wake-up approaches. Since the research efforts on the on-demand wake-up approach are very limited for the underwater environment, there are still many open issues. Therefore, this paper also presents several future research directions that need in-depth investigations to realize energy-efficient IoUT networks for various underwater applications.    

\section*{Acknowledgement}
 Figure 1 was created by Olga Kasimova, scientific illustrator at King Abdullah University of Science and Technology (KAUST).
 
 \bibliographystyle{IEEEtran}
 \vspace{-0.5 cm}
 \bibliography{ref}
 
 \renewenvironment{IEEEbiography}[1]
  {\IEEEbiographynophoto{#1}}
  {\endIEEEbiographynophoto}
  \vspace{-1.5 cm}
 \begin{IEEEbiographynophoto}
 {Muhammad Muzzammil} received the DoE degree in Information and Communication Engineering from Harbin Engineering University, China in 2021. His research interests lie in the areas of wireless communications, magneto-inductive communication and underwater acoustic communication and networking.
 \end{IEEEbiographynophoto}  
\vspace{-1.5 cm}
 \begin{IEEEbiographynophoto}
 {Nour Kouzayha} (Member, IEEE) received the Ph.D. degree in electrical and computer engineering from American University of Beirut (AUB), Beirut, in January 2018.  
 She is currently a Postdoctoral Fellow with the Information Theory Laboratory (ISL), King Abdullah University of Science and Technology, Saudi Arabia. Her research interests are in the area of wireless communications, stochastic geometry, Internet of Things, UAV networks, and THz communications.
 \end{IEEEbiographynophoto} 
 \vspace{-1.5 cm}
 \begin{IEEEbiographynophoto}
 {Nasir Saeed} (Senior Member, IEEE) received Ph.D. in Electronics and Communication Engineering from Hanyang University, Seoul, South Korea, in 2015. He is currently an Associate Professor with the Department of Electrical Engineering, Northern Border University, Arar, Saudi Arabia. His current research interests include underwater wireless communications, aerial networks, Industrial IoT, and localization.
 \end{IEEEbiographynophoto}
 \vspace{-1.5 cm}
  \begin{IEEEbiographynophoto}
 {Tareq Y. Al-Naffouri} (Senior Member, IEEE) received the Ph.D. degree in electrical engineering from Stanford University, Stanford, CA, USA, in 2004. He is currently a Professor with the Electrical Engineering Department, King Abdullah University of Science and Technology. His research interests lie in the areas of sparse, adaptive, and statistical signal processing and their applications, localization, machine learning, and network information theory.
 
 \end{IEEEbiographynophoto}

\end{document}